\documentclass[dvips]{article}

\usepackage{icrc2011}

\title{New improved Sum-Trigger system for the MAGIC telescopes}
\newcommand{\etal}{\MakeLowercase{\textit{et al. }}} % "et al."
\shorttitle{D. Haefner \etal New Sum-Trigger system for MAGIC}

\authors{Dennis Haefner$^{1}$, Thomas Schweizer$^{1}$, Francesco Dazzi$^{2,3}$, Daniele Corti$^{2}$}
\afiliations{$^1$Max-Planck-Institute for Physics, Munich, Germany\\$^2$INFN Padova, Italy\\$^3$University of
Udine, Italy}
\email{dhaefner@mpp.mpg.de}

\abstract{In 2007 a prototype of a new analog Sum-Trigger was installed in the MAGIC I telescope, which lowered the trigger threshold from 55 GeV to 25 GeV and led to the detection of pulsed gamma radiation from the Crab pulsar.
To eliminate the need for manual tuning and maintenance demanded by that first prototype, a new setup with fully automatic calibration was designed recently. The key element of the new circuit is a novel, continuously variable analog delay line that enables the temporal equalization of the signals from the camera photo sensors, which is crucial for the efficient detection of low-energy showers.
A further improvement is the much larger trigger area consisting of a fully revised configuration of overlapping summing patches.
The new system will be installed on both telescopes, MAGIC I and II, enabling stereo observation in Sum-Trigger mode. This will significantly improve the sensitivity in the very low energy regime of 20 to 100 GeV, which is essential in particular for detailed pulsar studies, as well as the observation of high-redshift AGNs and distant GRB events.
Here we like to present the results of functionality tests of a fully working prototype and the basic design of the final system.}

\begin{document}
\maketitle

%Begin the section.
\section{Introduction}

The two MAGIC telescopes on the Canary Island of La Palma are the largest IACTs worldwide and currently provide the lowest energy threshold of 55 GeV. A low threshold is essential for the observation of distant gamma-ray sources like high-redshift AGNs or GRBs. Due to the absorption of high energetic gamma-rays by the extragalactic background light the universe is less opaque for lower energy cosmic rays. Additionally, some sources like pulsars have a very steep gamma-ray spectrum above 10 GeV, whose profound study depends on a low energy threshold as well.
To distinguish between cosmic events and noise triggers from night sky background light, in particular in this very low energy regime, special efforts concerning the trigger system are necessary.
Besides a standard digital trigger with a threshold of 55 GeV, a prototype analog Sum-Trigger has been installed in October 2007 that lowered the trigger threshold significantly down to 25 GeV \cite{lab1,lab2}.
In principle this analog trigger system amplifies and sums up signals of patches of adjacent camera pixels using analog electronics. The main difference to other systems is that a discriminator is applied to the analog sum of pixels and not to each individual pixel. Having achieved excellent results in the detection of the pulsed gamma radiation from the Crab pulsar \cite{lab3} the analog Sum-Trigger now is being further optimized.

\section{New Sum-Trigger system}

 \begin{figure*}[th]
  \centering
  \includegraphics[width=15cm]{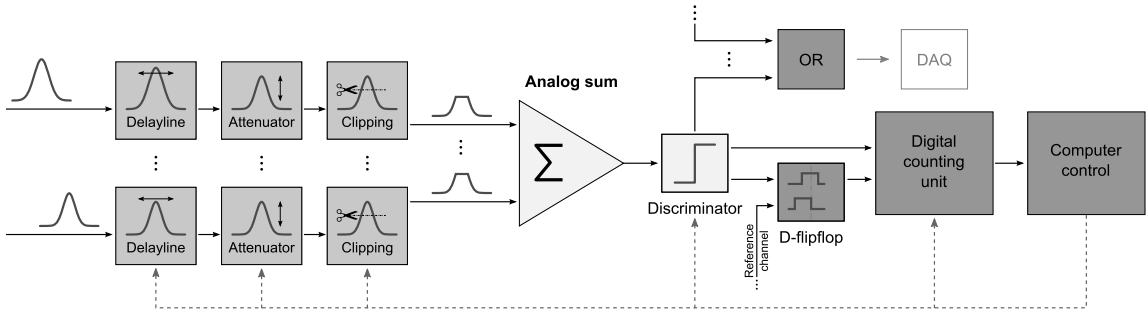}
  \caption{Sum-Trigger II principle schematic. On the Clip boards (medium gray shaded boxes) the incoming analog signals are adjusted in time and amplitude and clipping is performed. In the sum stage (light gray) the signals of one macrocell are summed up and discriminated. From there, the digital signal is fed into the ASTRO board (dark gray) on which the measurement circuits and computer control are located.}
  \label{fig_schematic}
 \end{figure*}

 \begin{figure*}[th]
   \centerline{\includegraphics[width=3.5cm]{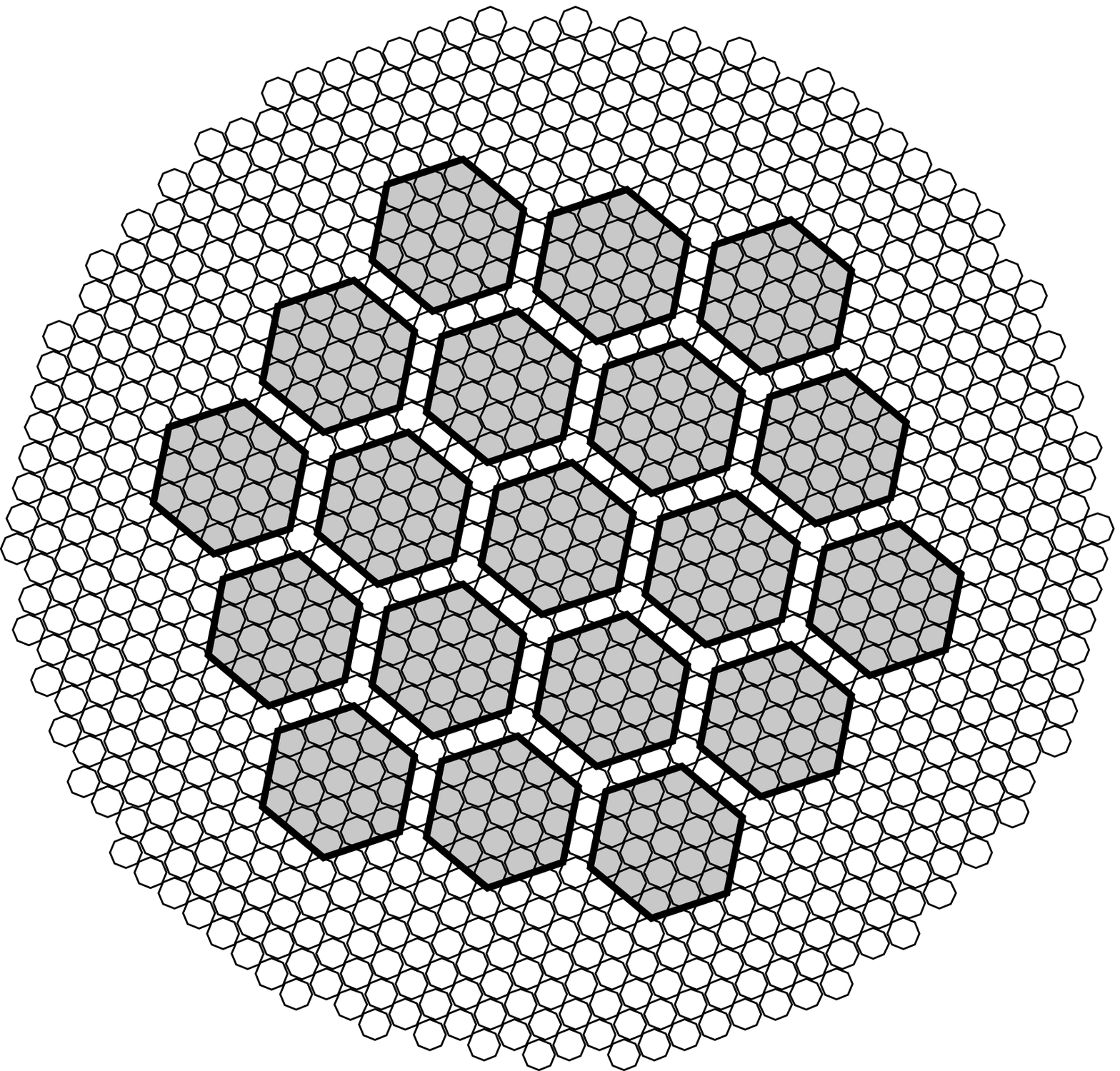}\label{fig1}
               \hfil
               \includegraphics[width=3.5cm]{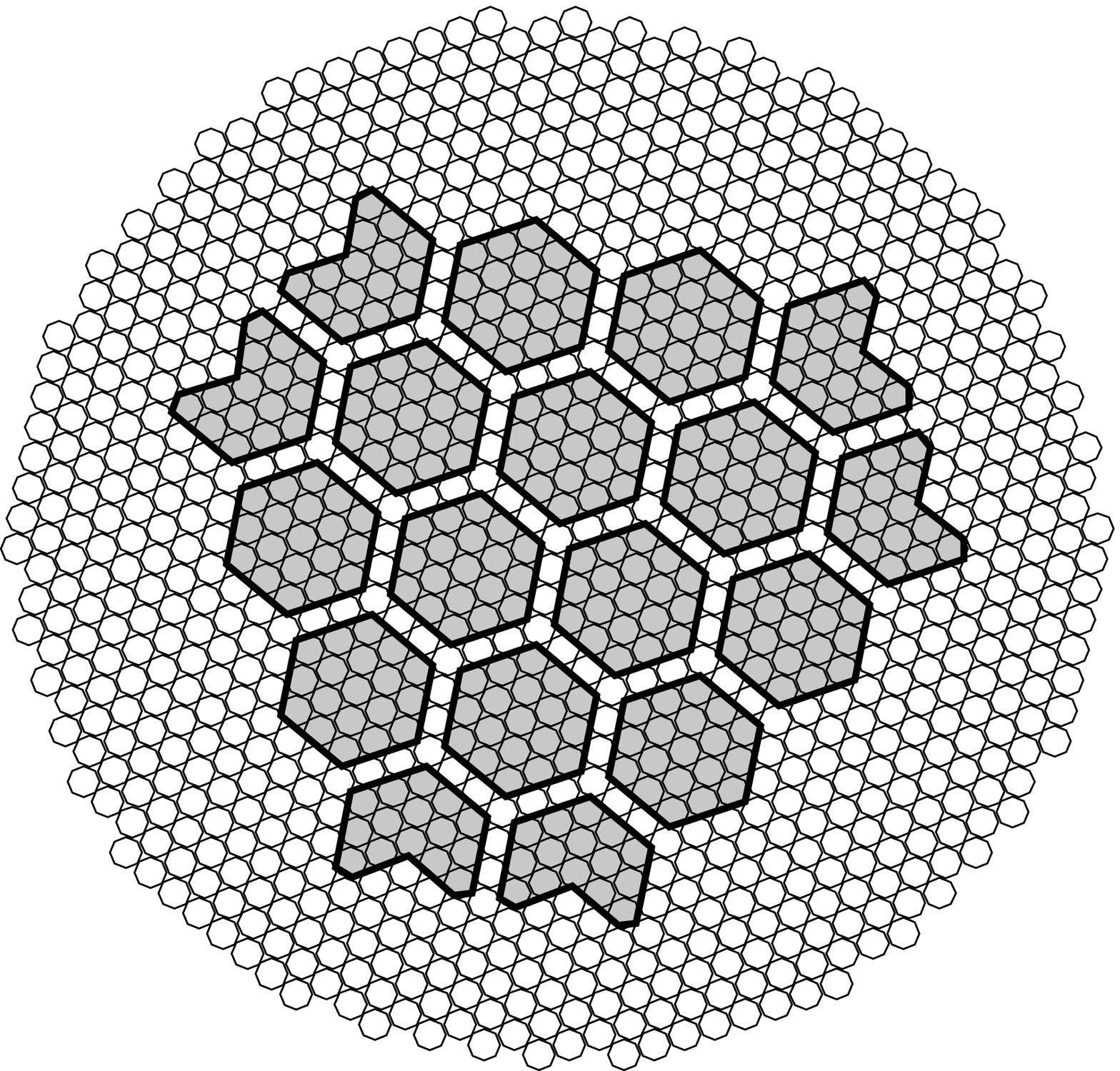}\label{fig2}
               \hfil
               \includegraphics[width=3.5cm]{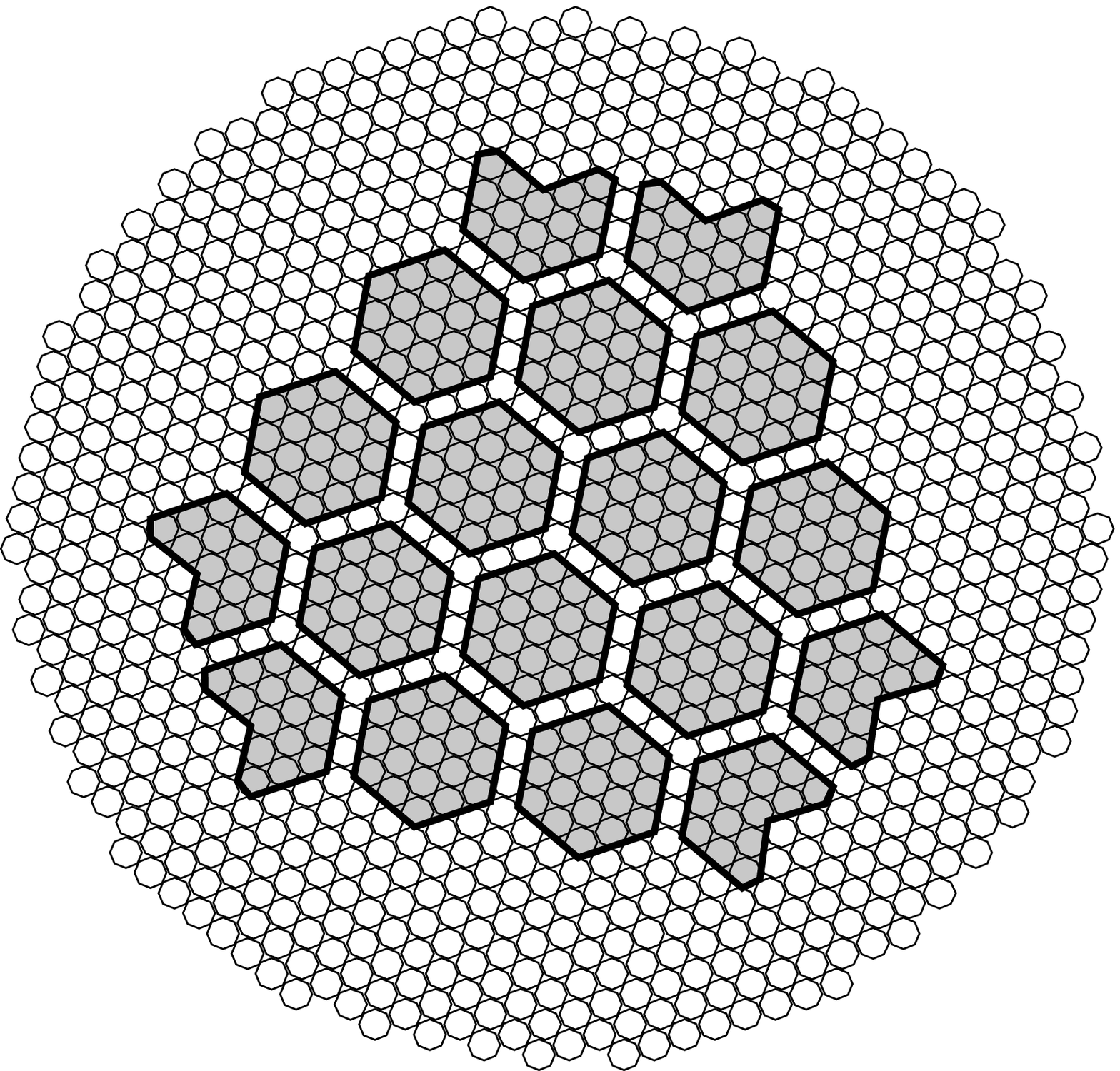}\label{fig3}
               \hfil
               \includegraphics[width=3.5cm]{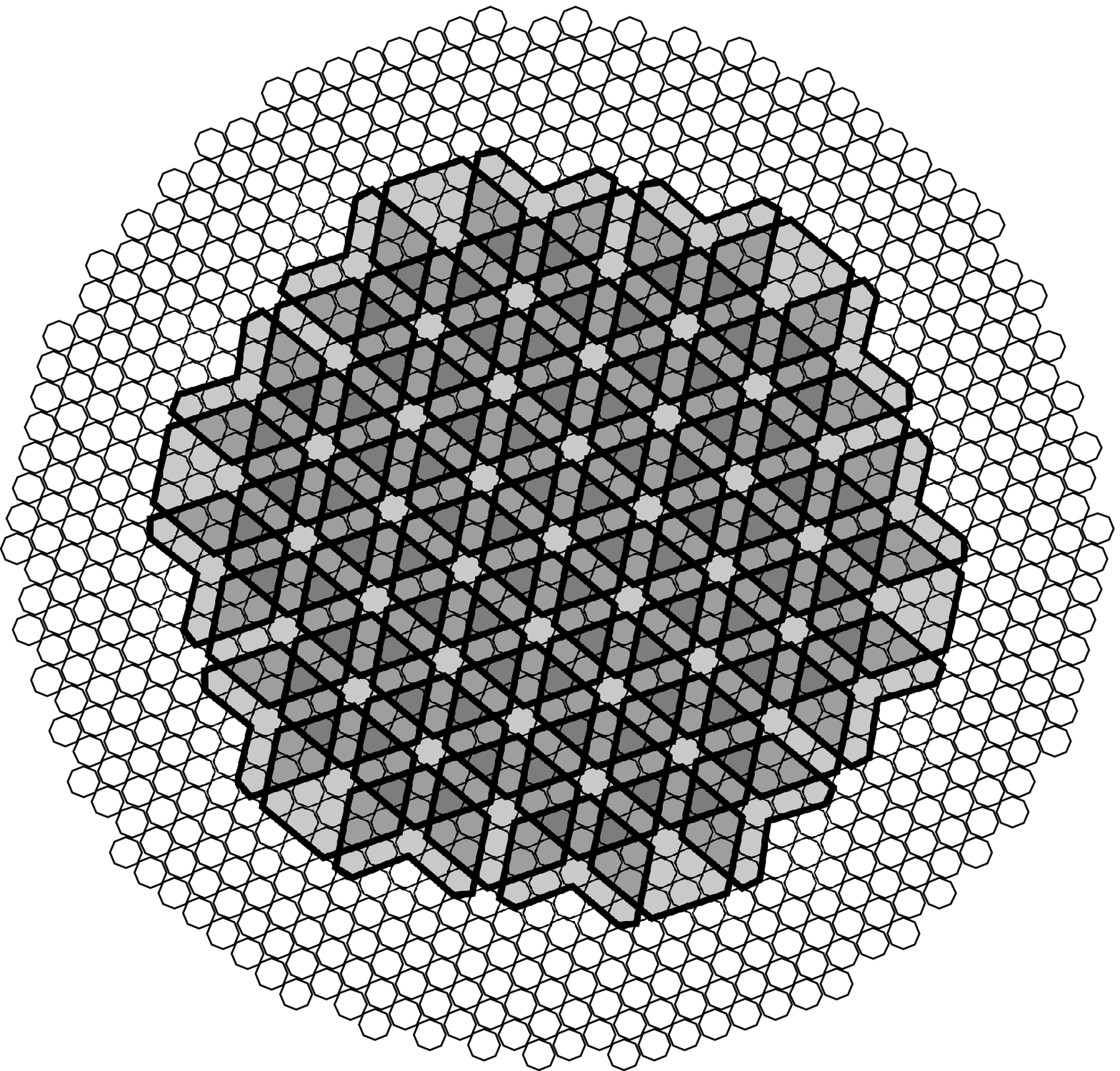}\label{fig4}
              }
   \caption{Schematic views onto the MAGIC II camera. Each pixels is represented by a small white hexagon. The gray shaded polygons depict the macrocells. The first three figures correspond to the first, second and third macrocell layer (left to right). The last figure shows all three layers superimposed.}
   \label{fig_mapping}
 \end{figure*}

\subsection{Overview}

Using the experience acquired during the construction and operation of the first prototype analog Sum-Trigger for MAGIC I, a new analog trigger, hereafter called Sum-Trigger II, is being designed. The main concept is to divide the inner part of the camera into several patches (macrocells), partly overlapping, in which the signals are summed up. The final trigger decision is derived from the summed signals of one patch. This method improves the signal to noise ratio\footnote{Here, the term \textit{signal to noise ratio} is used to describe the trigger sensitivity to very weak signals ``hidden'' in the noise floor.}, since all pixels within a macrocell contribute to the trigger decision, even if they are individually dominated by NSB fluctuations. A fundamental requirement is to preserve the features of the fast analog signal of about 2.6 ns FWHM from the camera's photomultiplier tubes (PMTs), in order to maintain the arrival time information of physical events and to minimize the time coincidence window of the sum and hence to reduce noise triggers. This implies electronics with high bandwidth and high signal integrity, very low skew between channels and a fast signal clipping stage on each channel to prevent triggers on large afterpulses caused by foreign atoms ionized and backscattered inside the PMTs \cite{lab4}.\\
Sum-Trigger II will be installed in both MAGIC telescopes and can be operated stand alone or in parallel with the current standard digital trigger, in single mode or in stereo mode included in the two-telescope coincidence trigger (level 3).\\
The main goals of the Sum-Trigger II development are the following:

\begin{itemize}
\item Enlarge the current Sum-Trigger area to the standard trigger's dimensions. This will allow observation in \textit{wobble mode} and increase the effective trigger area.
\item Reduce system dimensions, in order to enable the installation of two Sum-Triggers close to the signal receivers in the counting house.
\item Include automatic amplitude and delay adjustment. In the first Sum-Trigger version, amplitude and delay of each channel had to be manually adjusted in a time consuming process. An automatic procedure guarantees to save time, to keep the system well calibrated and to improve the low-energy performance.
\end{itemize}

Sum-Trigger II consists of four main subsystems (see also figure \ref{fig_schematic}):
\begin{itemize}
\item On the \textit{Clip Boards} the timing and gain of each channel is equalized and the analog signals are clipped.
\item The so-called \textit{Sum Backplane} routes the analog signals from the Clip Boards to the corresponding sum stages.
\item The \textit{Sum Boards} sum up the analog signals of the channels inside one macrocell and apply a discriminator circuit to the sum.
\item One \textit{ASTRO Board} incorporates the timing and amplitude measurement circuits and the computer control, as well as an OR-logic subsuming the trigger signals from all Sum Boards.
\end{itemize}

\subsection{New macrocell mapping}

Monte Carlo studies showed that Sum-Trigger II performance can be optimized in the range of 10 - 30 GeV for macrocells composed by around 20 pixels. Considering that the hexagonal shape guarantees both a symmetrical overlap and a central symmetry, a single patch is defined as a hexagon of 19 pixels, which is smaller than the standard trigger's macrocell (37 pixels).\\
The idea is to maintain a circular symmetry in which each macrocell is surrounded by six other macrocells, equally disposed every 60 degrees. The mapping is composed by 3 layers, two out of three with the same shape, but a different rotation (see figure \ref{fig_mapping}).\\
For the mapping, the analog signals of all pixels inside each macrocell have to be routed to their dedicated sum stage. This is performed on the so-called Sum Backplane, which is a completely passive circuit board with strict electronic constraints, in order to preserve the synchronicity among channels within 100 ps, keep the bandwidth above 1 GHz, and minimize cross-talk ($<$ 1\%).

\subsection{Adjustable analog delay line}

 \begin{figure}[!t]
  %\vspace{5mm}
  \centering
  \includegraphics[width=7cm]{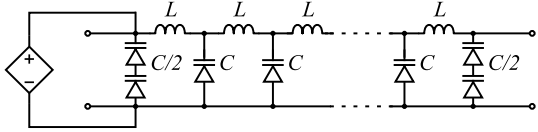}
  \caption{Simplified schematic of the delay line. Each stage consists of an inductor \textit{L} and a variable capacitor \textit{C}. The line is terminated with \textit{C/2} (two varactors in series) on both ends, in order to reduce frequency dependency.}
  \label{fig_delaylinecircuit}
 \end{figure}

 \begin{figure}[!t]
  %\vspace{5mm}
  \centering
  \includegraphics[width=8.25cm]{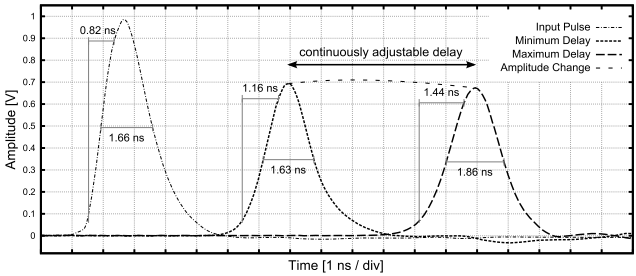}
  \caption{Measurements on the adjustable delay line module. Shown are the input pulse of about 1.7 ns FWHM and the output pulses for minimum and maximum delay.}
  \label{fig_delaylinepulses}
 \end{figure}

The most crucial component of the new Sum-Trigger is a newly developed, adjustable analog signal delay line that can compensate the different signal transition times of the PMTs and small delay differences of the 162m long optical fibers used for the analog transmission of the fast signals.
The delay line consists of 25 stages of second order passive low-pass filter circuits (figure \ref{fig_delaylinecircuit}) made of chip inductors and variable capacity diodes (\textit{varactors}). The adjustable capacities enable the control of the phase shift of each stage - and thus the overall delay of the whole chain of filter circuits - by applying a reverse voltage to the diodes.
On the current delay line a bandwidth of more than 420 MHz for minimum delay is achieved. The inductors and varactors were chosen such that the total delay range spans about 6 ns (see figure \ref{fig_delaylinepulses}).\\
With this delay line the timing differences of pulses coming from different pixels can be accurately compensated, which is essential for an optimal performance of the Sum-Trigger.

\subsection{Calibration process}

 \begin{figure}[!t]
  %\vspace{5mm}
  \centering
  \includegraphics[width=7cm]{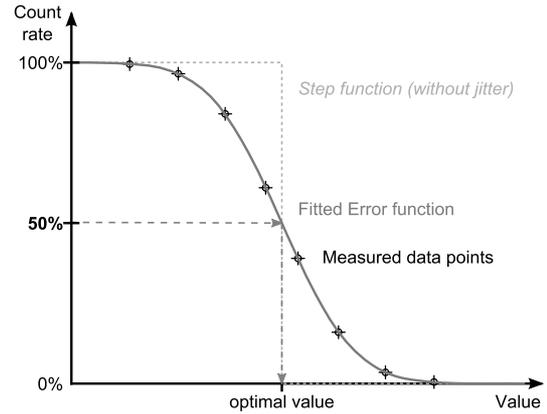}
  \caption{Principle of the measurement process. An Error function is fitted to the measured rates, revealing the optimal setting at a rate of 50\%. Here, ``value'' corresponds either to \textit{gain} or \textit{delay}, depending on which channel property is calibrated.}
  \label{fig_ratecountprinciple}
 \end{figure}

Due to occasionally required tuning of the high-voltages applied to the PMTs, the amplitude and signal transit time can change for individual pixels and have to be re-adjusted on a regular basis for optimal trigger performance. Hence, Sum-Trigger II will include a computer controlled automatic adjustment of amplitude and delay of the signals.
In order to keep the complexity of the new circuits low, an innovative measuring technique based on the evaluation of a series of rate measurements is introduced, requiring only very few additional electronics.
In particular, the discrete trigger output of the discriminator is used to measure amplitudes by counting the number of events that surpass the discriminator threshold within a certain time span.
Similarly, the rates to determine the delay of each channel are derived from the output of a D-type flip-flop that is used to compare the arrival times of two pulses.\\
To produce adequate reference signals, calibration LEDs, located in the center of the telescope reflector, emit light pulses at a well defined frequency and amplitude. Since the light pulses hit all camera pixels simultaneously with an equal intensity, they enable a relative timing and gain adjustment among all channels.\\
The distinctive feature of the new measuring technique is to take advantage of intrinsic jitter in amplitude and arrival time of the signals, to efficiently derive the optimal settings for the trigger components.
Relative amplitude variations are in the order of 10\%, and signal transit time jitters typically below 1 ns. These fluctuations primarily originate from the PMTs.\\
When performing the gain adjustment, the discriminator threshold is fixed to the target amplitude level, and the attenuation value is varied while counting the number of trigger signals from the discriminator. Likewise, the optimal delay is derived by tuning the delay line module. Here, the counter is incremented by the D-type flip-flop comparing the signal arrival time with the timing reference channel. The result is a series of rate measurements of the transition region from maximum to minimum number of trigger counts.
Due to the time and amplitude jitter inherent in the signals, the rate scans show a cumulative distribution function, which is used to derive the optimal settings, being found at 50\% of the maximum rate (figure \ref{fig_ratecountprinciple}).

\section{Prototype}

In order to verify the performance of the new concepts of automatic calibration and to examine new components such as the adjustable delay line, a small Sum-Trigger II prototype has been designed and temporarily installed at the MAGIC I telescope.
Furthermore, it was the basis for the development of most of the new techniques being applied in Sum-Trigger II.

\subsection{Prototype setup}

The prototype comprises all essential components of the final Sum-Trigger II, such as delay and gain adjustment, signal clipping, sum stage, the digital counting circuits and computer control (see figure \ref{fig_schematic}).
It was designed for a test patch of only 8 pixels.
With the prototype the performance of the automatic calibration has been be tested in detail, as well as the functionality of the delay line, the clipping amplifier and the computer control logic.

\subsection{Performance of the calibration process}

 \begin{figure}[!t]
  %\vspace{5mm}
  \centering
  \includegraphics[width=8.25cm]{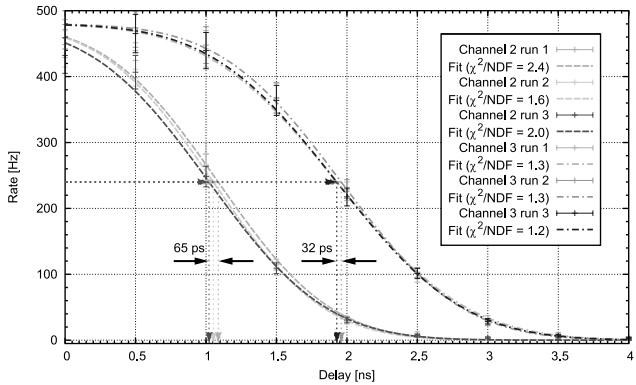}
  \caption{Displayed are cumulative distribution functions (here: \textit{Error functions}) fitted to data points recorded during delay calibration. Three consecutive calibration runs were performed on two channels, to test the stability of the procedure.}
  \label{fig_delayratescan}
 \end{figure}

 \begin{figure}[!t]
  %\vspace{5mm}
  \centering
  \includegraphics[width=6cm]{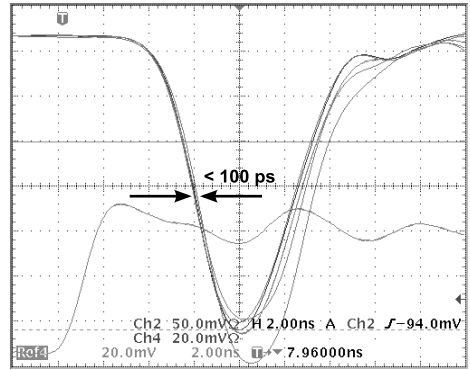}
  \caption{Oscilloscope image of pulses recorded on five different channels after delay calibration. Clearly visible is the coincidence of the rising edges of the (inverted) signals (arrows) within 100 ps or less.}
  \label{fig_risingedges}
 \end{figure}

The precision of the above described automatic calibration process is more than sufficient. In case of the delay adjustment, the variance in timing of all calibrated channels is significantly below 100 ps (figure \ref{fig_risingedges}) and the variation of the optimal delay value among repeated measurements on the same channel stays even below 70 ps (figure \ref{fig_delayratescan}). However, according to Monte Carlo simulations, an uncertainty of up to 250 ps is acceptable, and hence the number of data points per measurement can be reduced, speeding up the calibration procedure.
Also the gain adjustment performs satisfactorily, even though there were technical problems during the tests that reduced the precision of the equalization of the pulse amplitudes to around 10\%. Nevertheless, even this 10\% uncertainty does not affect the performance of the Sum-Trigger.\\
Conclusively, one can state that the new calibration procedure features excellent performance with only a minimum of additional electronics and complexity.
Including this functionality in the Sum-Trigger II will significantly improve Sum-Trigger's long-term stability and reliability and will help to minimize maintenance. The more precise temporal superposition of signals in the sum may further reduce the energy threshold below 25 GeV, though this has not been verified yet.

%\vspace{\baselineskip}

\clearpage

\end{document}